\documentclass{sig-alternate}

\usepackage{subfig}
\usepackage{graphicx}

\begin{document}


%
%
%
%
%

\title{Profiling-Assisted Decoupled Access-Execute} 
%
%
%
%
%

\numberofauthors{5} 
\author{
\alignauthor Jonatan Waern \\
       \affaddr{Uppsala University}\\
       \email{Jonatan.Waern.5948\\@student.uu.se}
\alignauthor Per Ekemark \\
       \affaddr{Uppsala University}\\
       \email{Per.Ekemark.8846\\@student.uu.se}
\alignauthor Konstantinos Koukos \\
       \affaddr{Uppsala University}\\
       \email{Konstantinos.Koukos\\@it.uu.se}
\and  
\alignauthor Stefanos Kaxiras\\
       \affaddr{Uppsala University}\\
       \email{stefanos.kaxiras\\@it.uu.se}
\alignauthor Alexandra Jimborean\\
      \affaddr{Uppsala University}\\
      \email{alexandra.jimborean\\@it.uu.se}
}

\maketitle
\begin{abstract}
 As energy efficiency became a critical factor in the embedded systems domain, dynamic voltage and frequency scaling (DVFS) techniques have emerged as means to control the system's power and energy efficiency. Additionally, due to the compact design, thermal issues become prominent.

State of the art work promotes software decoupled access-execution (DAE) that statically generates code amenable to DVFS techniques. The compiler builds memory-bound access phases, designed to prefetch data in the cache at low frequency, and compute-bound phases, that consume the data and perform computations at high frequency.

This work investigates techniques to find the optimal balance between lightweight and efficient access phases. A profiling step guides the \textit{selection} of loads to be prefetched in the access phase. For applications whose behavior vary significantly with respect to the input data, the profiling can be performed online, accompanied by just-in-time compilation. We evaluated the benefits in energy efficiency and performance for both static and dynamic code generation and showed that precise prefetching of critical loads can result in 20\% energy improvements, on average. DAE is particularly beneficial for embedded systems as by alternating access phases (executed at low frequency) and execute phases (at high frequency) DAE proactively reduces the temperature and therefore prevents thermal emergencies.
\end{abstract}

%
%

%



\section{Introduction}

Embedded systems evolved dramatically since the early designs, with a significant rise in processing power and functionality, providing nowadays a wide range of functionalities, from household electronics to complex high performance consoles and hand-held devices. To satisfy the increasing demands, the architecture of emerging embedded systems includes multiprocessors and multi-cores, endowed with cache memories and sophisticated hardware, resembling more and more the traditional computer systems.

Power and energy efficiency became a critical factor in hardware design, not only as a limiting factor for performance, but also increasingly important for battery lifetime, given the prevalence of mobile devices. Moreover, the reduced size of embedded systems combined with growing demands for high-performance, increases power density and adds another critical factor, namely temperature and its inherent impact on performance, reliability, power consumption and cooling costs.

A significant body of work tackled energy efficiency using DVFS techniques and  scaling down voltage  for energy management~\cite{Andrei02,Jejurikar04,Liao06,Martin02}. However,  such techniques are no longer applicable with the break-down of Dennard scaling~\cite{MOSFET}. 

To address this problem, we have previously proposed a software decoupled access-execute (DAE) scheme in which the compiler generates code that yields high performance and is better suited for DVFS techniques~\cite{Jimborean14}.  DAE is based upon the idea of generating program slices~\cite{Weiser81} with respect to read memory accesses, creating one memory-bound phase (Access phase) that prefetches data into cache, and a compute-bound phase (Execute phase) that consumes the prefetched data for program execution. 
DAE exploits the fact that reducing frequency during memory-bound phases saves energy without harming performance, while automatically generated coarse phases reduces the number of time frequency is scaled. Running the processor at a low frequency during the Access phase and at a high frequency during the Execute phase leveraged Energy Delay Product (EDP) improvements of 25\% on average on statically analyzable codes. Nevertheless, the heuristics employed in generating lightweight and efficient access phases for scientific and task-based parallel codes are not applicable on applications with complex control-flow and irregular memory accesses.

This proposal departs from DAE for task-based parallel codes~\cite{Jimborean14} and provides methods to  efficiently prefetch long latency loads with hard to predict access patterns. Our approach combines profiling, instruction reordering and just-in-time compilation to generate Just-in-time decoupled access execute phases. Guided by profiling, PDAE is precise in targeting the critical loads and rips the benefits of timely software prefetching. Using the program slicing approach to generate access phases (as opposed to static heuristics to simplify the control-flow and target indirect memory accesses employed in  DAE~\cite{Jimborean14}), PDAE contains the required control instructions to reach all critical loads. We have evaluated the performance of statically generated access phases and just-in-time compiled phases, underlining the overhead of JIT-ing. We recommend using JIT only for applications whose critical loads are input dependent and for cross-compiled applications (in which case the critical loads may vary with respect to the underlying architecture). Otherwise, an offline profiling step is sufficient to identify long latency loads and generate efficient access phases with no additional cost.

PDAE brings (1) energy efficiency, by scaling down frequency during memory bound phases;  (2) with a minimal impact on performance (5\% slowdown on average, and up to 20\% speed-up on memory bound codes); and (3) improvements in the system's thermal profile, by alternating high- and low-frequency execution phases, which naturally regulates the chip's temperature. Overall, this yields higher energy efficiency and performance and leads to increased reliability and diminished cooling costs.
%


\begin{figure}
\centering
\includegraphics[width=\linewidth]{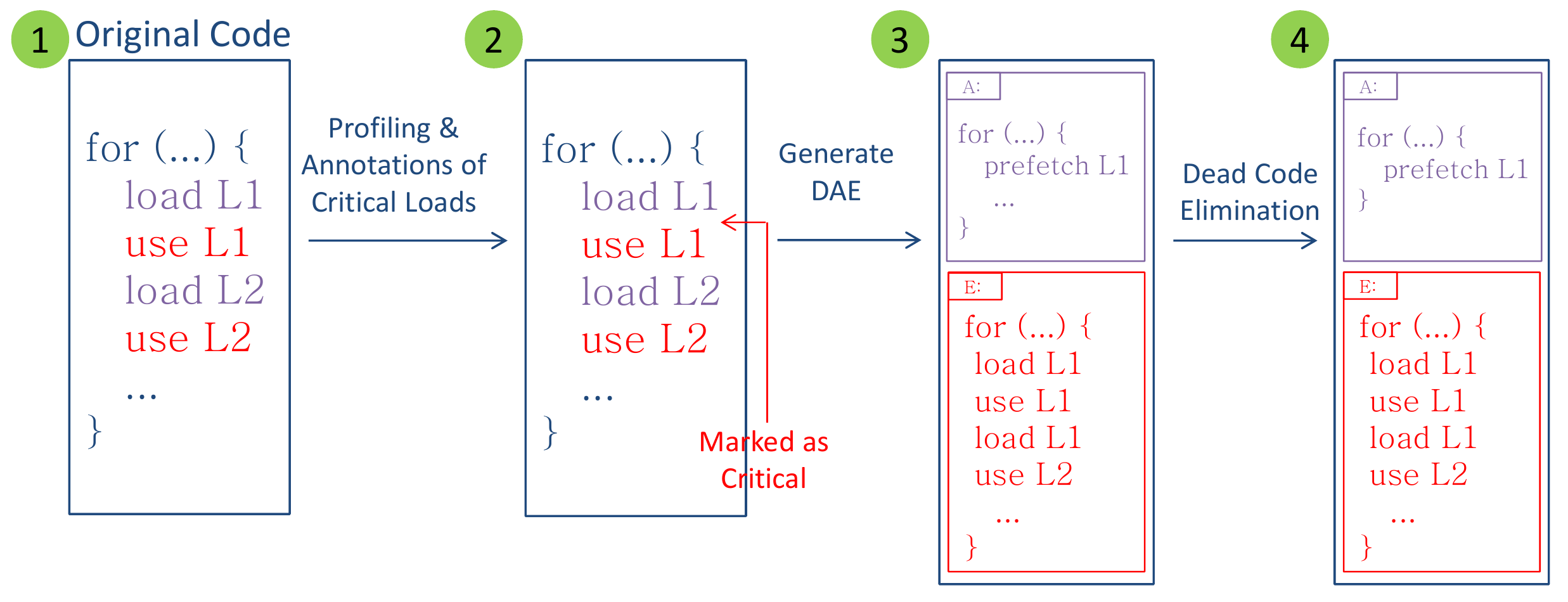}
\caption{Offline profiling and static generation of decoupled access-execute phases.}
\label{fig:static}
\end{figure}

\section{Methodology}

PDAE targets loops, which are executed in \textit{slices}. The decoupled access-execute scheme is applied to each loop-slice to ensure that the data prefetched during the access phase is consumed by the execute phase before being evicted from the cache. The access phase prefetches a selection of loads, while the execute version represents the original loop code, which now became compute-bound since the required data is available in the cache. The compiler passes are implemented in the LLVM compilation framework~\cite{LLVM}.

\subsection{Statically generated Access versions}
First, an offline profiling step is performed to identify long latency loads.  The compiler then proceeds as follows: (1) slices the loops and (2) generates Access - Execute versions for each loop-slice. Access phases contain the control-flow instructions and the computation of addresses in order to prefetch the target addresses of the critical loads. The instructions of the access phase are identified following the use-def chain. Hence, the access version is derived from the original code, but includes only a subset of its instructions. The process is detailed in Figure~\ref{fig:static}.

\begin{figure*}
\centering
\includegraphics[width=\textwidth]{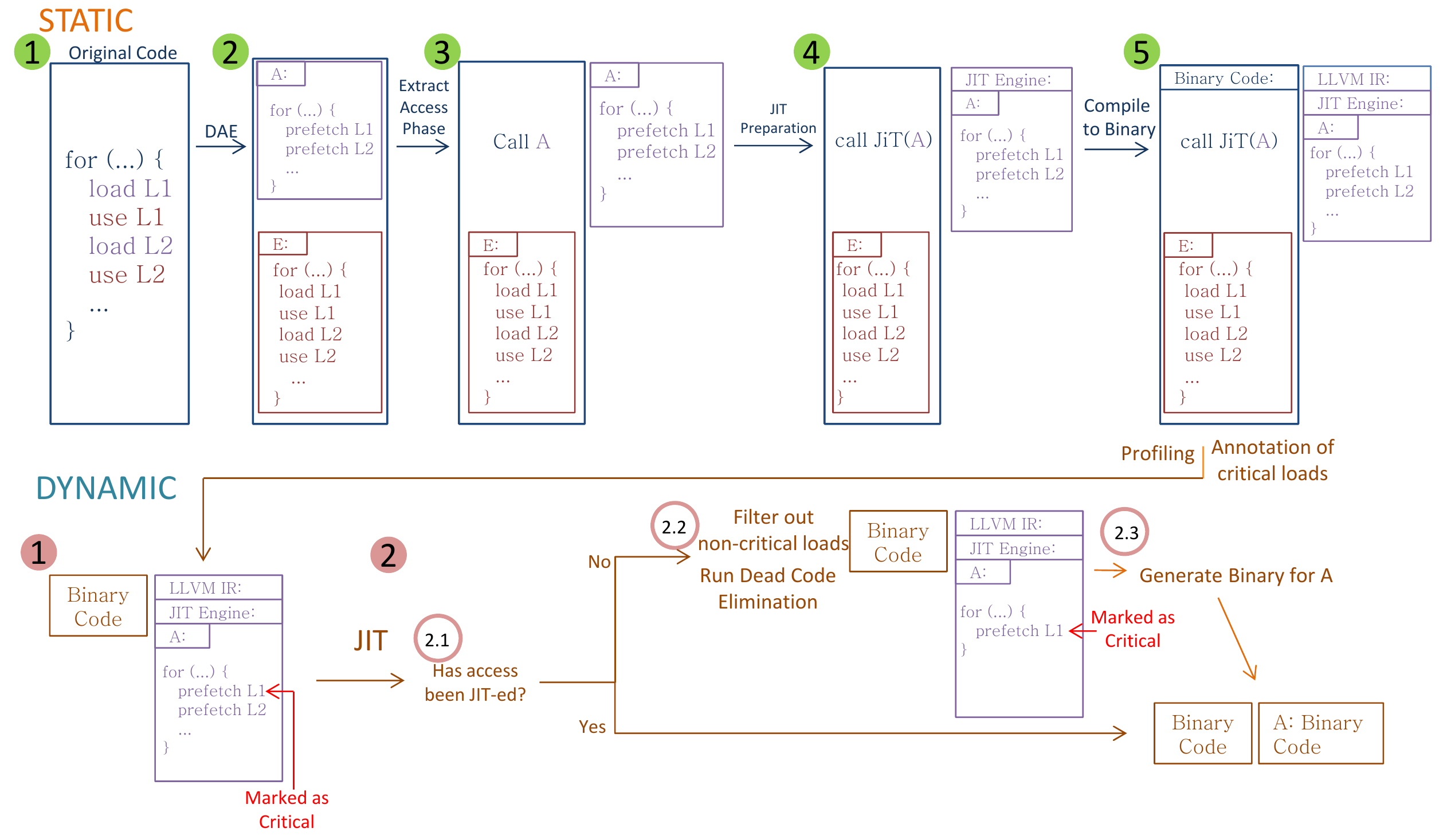}
\caption{Static and dynamic compilation steps for generating profiling-assisted DAE with just-in-time compilation.}
\label{fig:dynamic}
\end{figure*}
%

\subsection{Dynamically generated Access versions}
Similarly, (1) the loop is first sliced, (2) then the compiler generates Access-Execute versions per loop slice, however, since no profiling information is yet available, all loads are prefetched in the base-Access phase. Next, (3) the access version is extracted in a new function and support for just-in-time (JIT) compilation is added. Only access phases are JIT-ed, while (4) the rest of the code is statically compiled, to reduce the overhead. 

At runtime, (1) a profiling step is used to identify critical loads and to map them to the corresponding prefetch instructions in the access version. The first invocation of the access phase will trigger (2) a JIT compilation, which will filter out prefetch instructions which do not map a critical load and will clean dead code, yielding a lean and efficient access phase. Future invocations of the access version will directly call the already generated access code. The static and dynamic compilation steps are shown in Figure~\ref{fig:dynamic}.

\begin{figure}
\centering
\includegraphics[width=\linewidth]{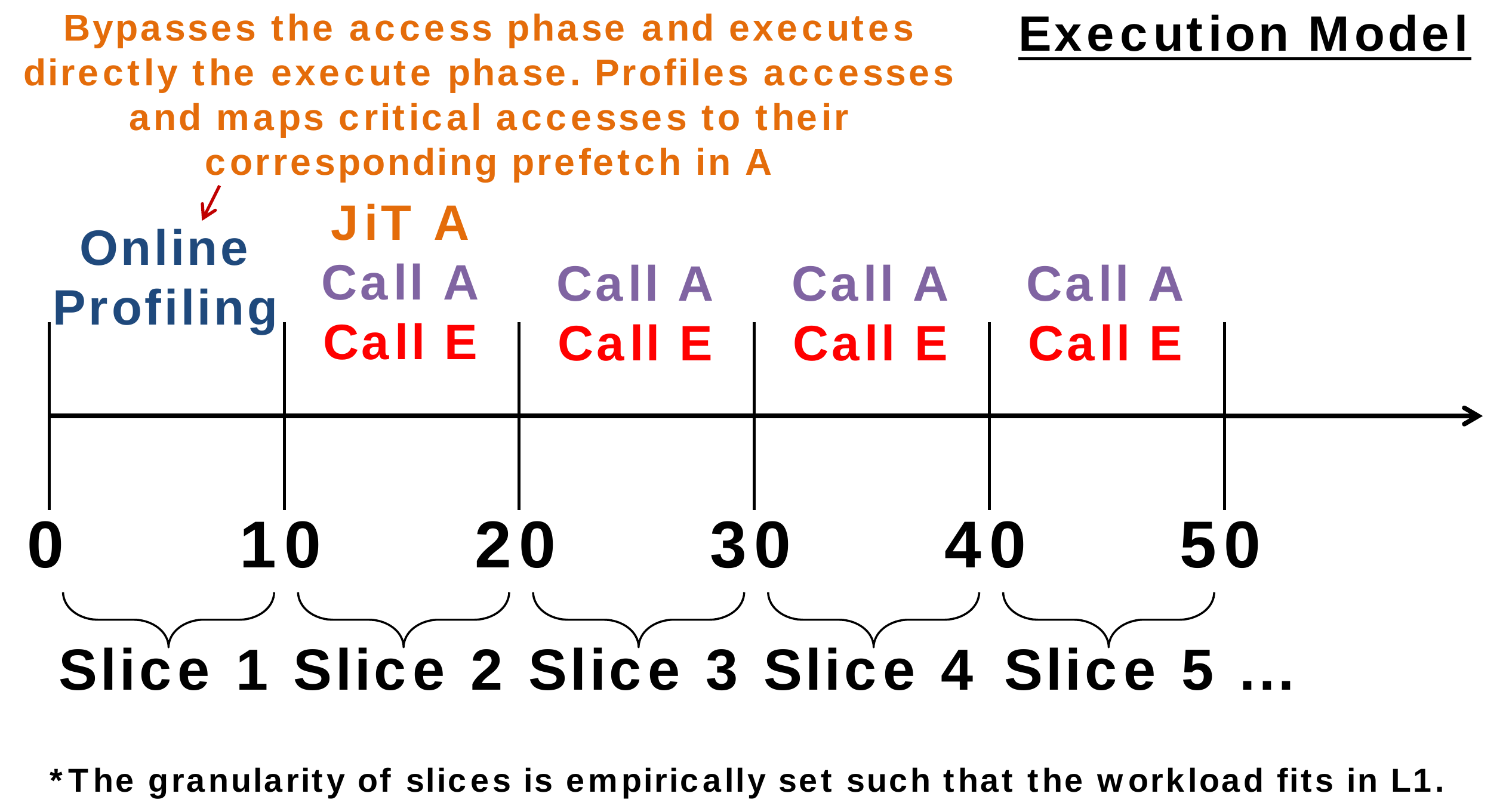}
\caption{Execution of each target loop consists of three steps: profiling, JIT compilation of the customized Access version and execution of Access-Execute pairs to complete the loop execution.}
\label{fig:execmodel}
\end{figure}

The execution model is shown in Figure~\ref{fig:execmodel}. To complete the first loop-slice, the original (execute) version of the loop is invoked and the long latency loads identified by profiling are mapped to the corresponding prefetch instructions in the access version. The second slice of the loop invokes the access version for the first time, triggering a JIT compilation, while the remaining slices will complete loop execution without further compilation overhead.

\subsubsection{Profiling}
The profiling can be performed statically or dynamically on one loop-slice or based on sampling to reduce the overhead. As low-overhead profiling was discussed in numerous previous work~\cite{Mittal87, Sembrant12,HirzelChilimbi2001}, this proposal does not attempt to design new profiling methods. For this work, the experiments were conducted using an offline profiling step to annotate critical loads~\footnote{Instructions are annotated with LLVM metadata.} (even for the JIT-ed version) and we leave for future work integration with state-of-the art low-overhead profiling techniques~\cite{HirzelChilimbi2001}, tailored to identify long latency loads.

\subsubsection{Generating decoupled access-execute code}
The compiler proceeds by generating a base-Access version which includes software prefetch instructions for all loads in the original code, their requirements, and instructions that maintain the control flow (branches and computations of conditions). 
The execute phase will then ``consume'' from the L1 cache, the data loaded in the access phase. To this end, the compiler (1) creates a clone of the target loop-slice to derive the access phase, (2) selects the loads and the instructions required to compute their addresses, as long as they do not modify globally visible variables, (3) keeps in the access phase (i.e., cloned version) only the list of selected instructions and the control flow instructions, and removes all other computations and (write) memory accesses, (4) simplifies the control flow to eliminate dead code (e.g., empty basic blocks and unnecessary branches) present in the access phase, (5) the execute version is merely the original, sliced loop version. The base-Access version is generated statically to reduce the overhead of dynamic code transformations. Furthermore, the access version is extracted in a new function using the LLVM-extract tool, enabling in this manner that only the access version is compiled dynamically,
rather than the entire application.

\subsubsection{ JIT optimizations}
To enable dynamic optimizations of the base-Access version, we rely on the LLVM MCJIT engine~\cite{mcjit}.  The program is compiled into a dynamic library loaded by the JIT engine and the calls to the access version are replaced with JIT callbacks, which enable optimizing and generating the access version on demand. JIT callbacks return a function pointer to the access version. 

At execution time, the JIT engine loads the code of the access version (in the LLVM intermediate  representation) on start-up, together with the dynamic library which contains the program and continues by calling the ``main'' function to start the execution of the program. Upon a callback, the JIT first checks whether the callback for this access version has been resolved previously. On the first invocation, the JIT finds the code in LLVM IR corresponding to the access version and runs
a pass to remove extraneous prefetches, namely prefetch instructions that do not carry metadata information which indicates that they map critical loads. A dead code elimination pass follows to remove any unnecessary address computations and branches (i.e. corresponding to the eliminated prefetch instructions). This leads a lightweight but efficient access version, which is then JIT compiled to an in-memory executable binary. The corresponding function pointer is stored within the JIT so that  future calls of the access version do not trigger a new dynamic compilation. Finally the JIT returns the function pointer to the generated access phase, and the program continues execution by calling that function, followed by the execute phase. 

\subsection{The Power Model}
We used the same power model~\cite{pows} as the one employed to measure the energy expenditure 
in previous DAE proposals~\cite{Jimborean14, koukos13}. The model approximates power usage based on the metric Instructions per Cycle (IPC) which is collected using the PAPI library~\cite{papi99}. 

\begin{figure*}[ht]
     \centering
     \subfloat[Normalized Execution Time]{{\includegraphics[width=0.475\textwidth]{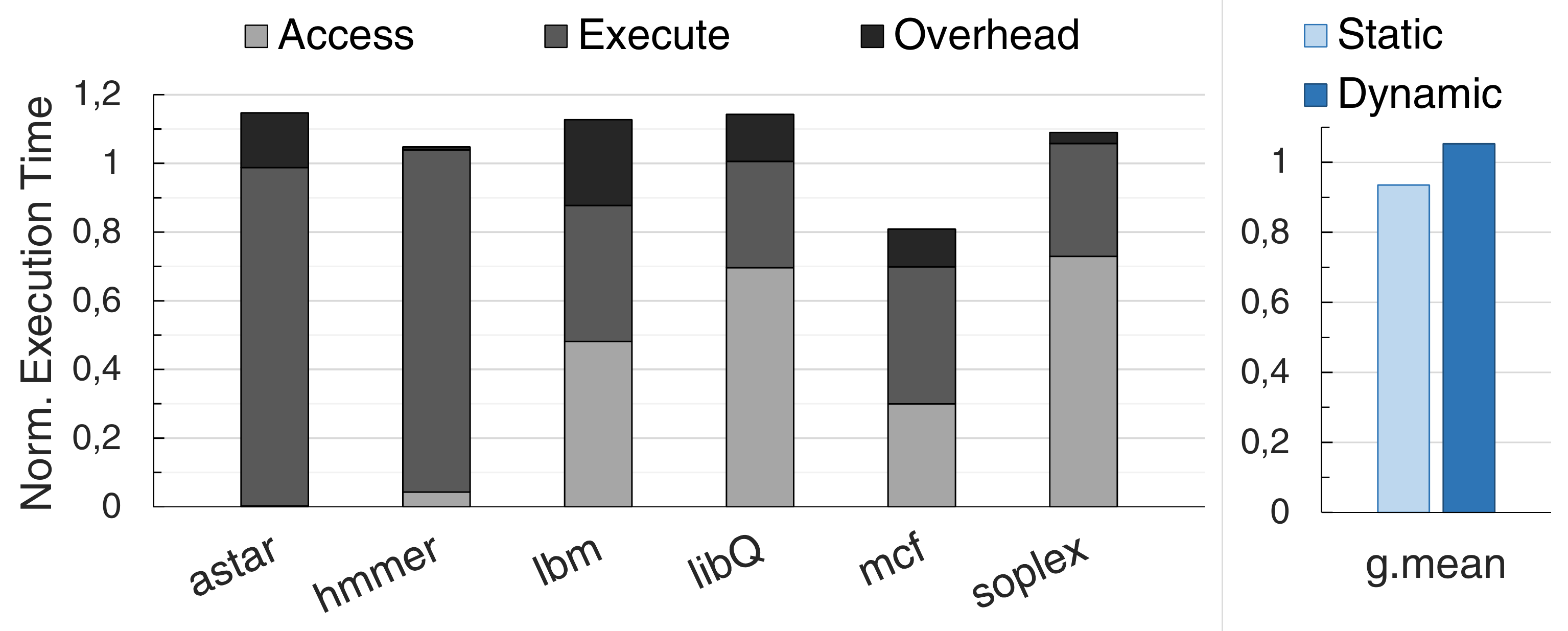} }}%
     \qquad
     \subfloat[Normalized Execution Energy]{{\includegraphics[width=0.475\textwidth]{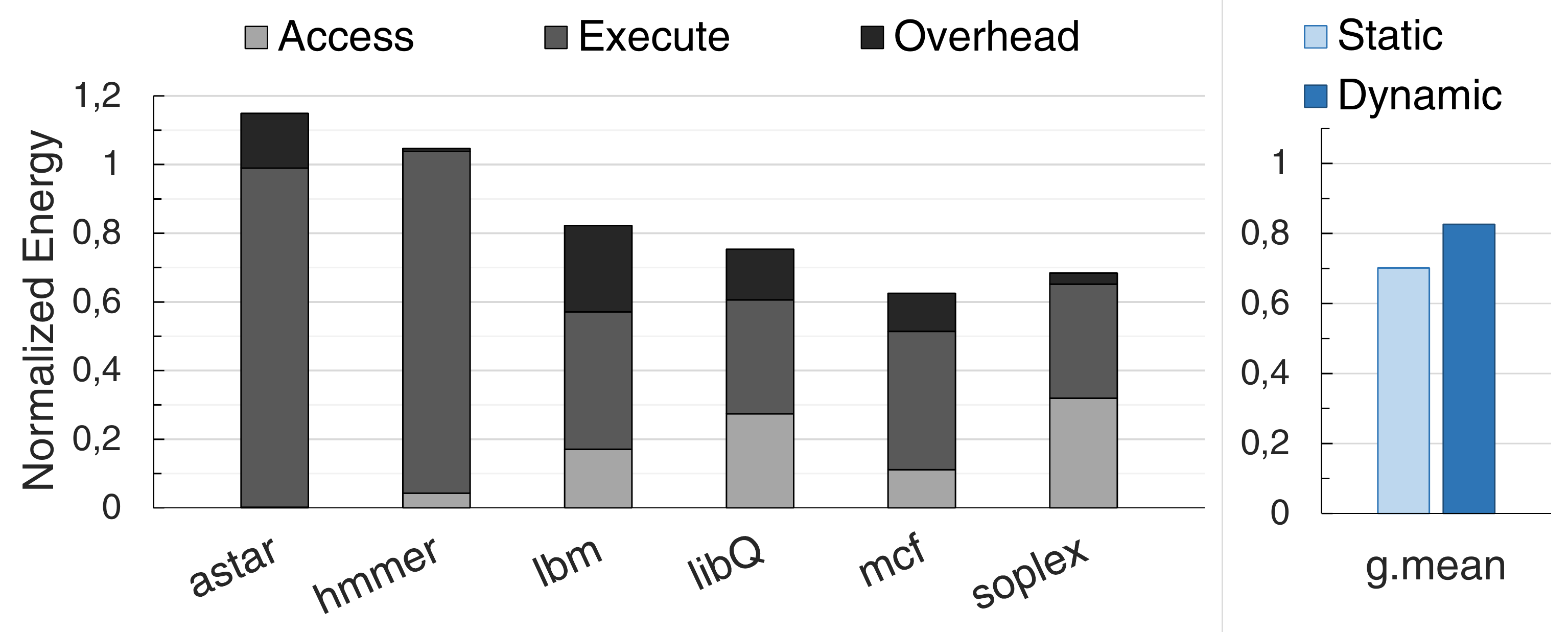} }}%
     \caption{Distribution of execution time and energy for a selection of SPEC 2006 benchmarks, on an Intel i7-2600K.}%
     \label{fig:Eval}%
 \end{figure*}
 
 \section{Evaluation}
 
We perform our measurements on an Intel Sandybridge i7-2600K processor with 16GBytes of DDR3 1333MHz memory. 
 
PDAE was evaluated on a subset of benchmarks from the  SPEC CPU-2006 benchmark suite ranging from compute bound to memory bound \textemdash for example, \emph{astar}
and \emph{hmmer} are compute bound while the rest are memory bound. For the evaluation we run 4 instances of the application, pinning one per core, to stress the memory bandwidth and emulate a more realistic execution (where system's shared
resources are highly utilized).

Figure~\ref{fig:Eval} shows the execution breakdown for each application. We measure time and energy separately for the \emph{Access} and \emph{Execute} phases.
For the static approach (using offline profiling and static compilation), the total time and energy per application are obtained by summing up the time  (energy, respectively) of the access and execute phases, as reported in Figure~\ref{fig:Eval}.

For the dynamic approach, we add the total \emph{Overhead}, including the JIT and the overhead of adjusting the frequency (considering nanosecond-scale DVFS). The overhead of online profiling is not considered in this evaluation, but is reported to be as low as  3-18\% by state of the art works~\cite{HirzelChilimbi2001}. Results are normalized to the original execution at maximum frequency. 

The static approach yields 7\% performance improvements on average across all applications and 25\% energy savings. For the dynamic approach, we observe a 5\% slowdown on average due to the JIT inherent overhead and an energy benefit of $\approx18\%$. 
As DAE exploits memory-bound phases to reduce energy,  compute-bound applications as \emph{astar}
and \emph{hmmer} cannot benefit since the access phase is insignificant.
For the memory bound applications we observe considerable energy savings of up to 35\% (for mcf). 

On average PDAE yields approximately 18\% energy savings for all benchmarks, while memory bound applications savings range within 20 -- 35\%. In terms of performance, we observe that a significant fraction of the execution time is spent in the access phase, which brings energy benefits. For \emph{mcf}, the  memory level parallelism (MLP) exposed in the access phase hides the
JIT overhead and results in 20\% performance speedup. For the rest of the applications we observe a slight slowdown due to the dynamic compilation, but since a significant fraction
of the execution time is spent in the access phase overall, significant energy savings are obtained.

\section{Related Work}
Dynamic voltage and frequency scaling techniques have been widely used for optimizing embedded syatems for power and energy efficiency~\cite{Andrei02,Jejurikar04,Liao06,Martin02,Liu07thermalvs,Bhatti11,Genser10,tempdvfs}. As thermal issues became a concern, some of these proposals focused on controlling the peak temperature. Liu et al~\cite{Liu07thermalvs} propose design-time DVFS planning and provide solutions to ensure optimal peak temperature, which may not correspond to the optimal energy solutions. Hence, the outcome is a thermal-constrained energy optimization. Bhatti et al~\cite{Bhatti11} study the interplay between state of the art Dynamic Power Management (DPM) and Dynamic Voltage \& Frequency scaling (DVFS) policies, and propose a scheme that selects at runtime the best-performing policy for any given workload. Genser et al~\cite{Genser10} acknowledged the importance of DVFS techniques for embedded systems and developed a power emulator to enable designers determine the most suitable voltage regulator and meet performance, power and energy demands.
While these approaches enhance the design of the embedded system, our focus is on adapting the applications to better suit the already available hardware.

Our compile-time code transformations to generate decoupled access-execute code build upon the decoupled access-execute for task based parallel systems~\cite{Jimborean14}. As this previous work was tailored for statically analyzable code, the compiler heuristics would be inefficient on irregular, pointer intensive, complex control flow applications. As a solution, we guide the generation of the access phase with a profiling step and evaluate both static and dynamic compilation approaches.
Previous static approaches that performed similar code transformations used inspector-executor methods~\cite{Salz91,CTY94,Rauchwerger95,Yokota02,Arenaz04} to monitor the behavior of the application during the \textit{inspecor} phase and optimize it during the \textit{executor} phase. Helper threads~\cite{Kamruzzaman11, mitosis05, Zhang07} were used to hide memory latency by prefetching data in advance, but, without the considering dynamic memory accessing behavior, for irregular codes such techniques might be generate overly complex and inefficient access phases.

While offline profiling and feedback driven compilation~\cite{Cooper05} has been shown to provide good results, dynamic compilation became increasingly popular \cite{javajit, adaptivejit, dynamo} to perform runtime optimizations.  Suganuma et al.\cite{javajit} implement a JIT compiler for Java, Arnold et al.\cite{adaptivejit} enable dynamic inlining using a JIT, while Bala et al~\cite{dynamo} use dynamic profiling to perform optimizations on-the-fly. The complexity of the runtime optimization and the size of the code to be compiled dynamically determine the JIT overhead. In PDAE we reduced this overhead to a minimum by preparing the base-Access phase statically and only removing unnecessary instructions on-the-fly. The small size of a loop slice also ensured that the JIT overhead is low. 

\section{Conclusions}
As DVFS techniques became popular for managing power and energy efficiency, we demonstrate a compiler method to transform applications to better suit the hardware DVFS capabilities. In particular, we target applications which contain irregular memory accesses and dynamic control flow and generate decoupled access-execute code versions. Accesses phases, dedicated to prefetch data to the cache, are generated automatically by the compiler, guided by profiling to ensure their efficiency. Being memory bound, access phases are run at low frequency, thus saving energy. Execute phases represent the original code, but become compute bound as they consume the data brought by the corresponding access phase to the cache. Execute phases are run at high frequency to maintain high-performance. Alternating access (low frequency) and execute (high frequency) phases naturally regulates the chip's temperature, with a positive side-effect on thermal related aspects (cooling costs, reliability, peak temperature, etc). We have evaluated both a static and a dynamic compilation approach, assisted by offline and online profiling respectively. PDAE assumes perfect profiling for identifying critical loads whose prefetch during the access phase provides energy benefits. The static scheme provides 25\% energy savings and 7\% performance improvement, compared to the original code run at maximum frequency. The dynamically compiled scheme adds the JIT-overhead and yields energy savings of 18\% on average (20-35\% for memory bound applications) with only 5\% performance degradation on average (up to 20\% \textit{speed-up} for memory bound applications).

\section{Acknowledgments}
This work is supported, in part, by the Swedish Research 
Council UPMARC Linnaeus Centre and the EU Project: 
LPGPU FP7-ICT-288653.

\bibliographystyle{abbrv}
\bibliography{references}  

\begin{thebibliography}{10}

\bibitem{mcjit}
{MCJIT} design and implementation.
\newblock http://llvm.org/docs/MCJITDesignAndImplementation.html.

\bibitem{Andrei02}
A.~Andrei, M.~Schmitz, P.~Eles, Z.~Peng, and B.~M. Al-Hashimi.
\newblock Overhead-conscious voltage selection for dynamic and leakage energy
  reduction of time-constrained systems.
\newblock In {\em Design, Automation and Test in Europe Conference and
  Exhibition}, pages 518--523, 2004.

\bibitem{Arenaz04}
M.~Arenaz, J.~Touriño, and R.~Doallo.
\newblock An inspector-executor algorithm for irregular assignment
  parallelization.
\newblock In J.~Cao, L.~Yang, M.~Guo, and F.~Lau, editors, {\em Parallel and
  Distributed Processing and Applications}, volume 3358 of {\em Lecture Notes
  in Computer Science}, pages 4--15. Springer Berlin Heidelberg, 2005.

\bibitem{adaptivejit}
M.~Arnold, S.~Fink, D.~Grove, M.~Hind, and P.~F. Sweeney.
\newblock Adaptive optimization in the jalapeno jvm.
\newblock {\em SIGPLAN Not.}, 35(10):47--65, 2000.

\bibitem{dynamo}
V.~Bala, E.~Duesterwald, and S.~Banerjia.
\newblock Dynamo: A transparent dynamic optimization system.
\newblock {\em SIGPLAN Not.}, 35:1--12, 2000.

\bibitem{tempdvfs}
M.~Bao, A.~Andrei, P.~Eles, and Z.~Peng.
\newblock Temperature-aware task mapping for energy optimization with dynamic
  voltage scaling.
\newblock In {\em Design and Diagnostics of Electronic Circuits and Systems
  (DDECS)}, 2008.

\bibitem{Bhatti11}
M.~K. Bhatti, C.~Belleudy, and M.~Auguin.
\newblock Hybrid power management in real time embedded systems: An interplay
  of dvfs and dpm techniques.
\newblock {\em Real-Time Systems}, 47(2):143--162, 2011.

\bibitem{CTY94}
D.~K. Chen, J.~Torrellas, and P.~C. Yew.
\newblock An efficient algorithm for the run-time parallelization of doacross
  loops.
\newblock In {\em Proceedings of the 1994 ACM/IEEE Conference on
  Supercomputing}, pages 518--527, 1994.

\bibitem{Cooper05}
K.~D. Cooper, A.~Grosul, T.~J. Harvey, S.~Reeves, D.~Subramanian, L.~Torczon,
  and T.~Waterman.
\newblock Acme: Adaptive compilation made efficient.
\newblock {\em SIGPLAN Not.}, 40(7):69--77, 2005.

\bibitem{MOSFET}
R.~Dennard, F.~Gaensslen, V.~Rideout, E.~Bassous, and A.~LeBlanc.
\newblock Design of ion-implanted mosfet's with very small physical dimensions.
\newblock {\em IEEE Journal of Solid-State Circuits}, 9(5):256 -- 268, 1974.

\bibitem{Genser10}
A.~Genser, C.~Bachmann, C.~Steger, R.~Weiss, and J.~Haid.
\newblock Power emulation based dvfs efficiency investigations for embedded
  systems.
\newblock In {\em International Symposium on System on Chip (SoC)}, pages
  173--178, 2010.

\bibitem{HirzelChilimbi2001}
M.~Hirzel and T.~Chilimbi.
\newblock Bursty tracing: A framework for low-overhead temporal profiling.
\newblock In {\em ACM Workshop on FeedbackDirected and Dynamic Optimization
  (FDD)}, 2001.

\bibitem{Jejurikar04}
R.~Jejurikar, C.~Pereira, and R.~Gupta.
\newblock Leakage aware dynamic voltage scaling for real-time embedded systems.
\newblock In {\em Design Automation Conf. (DAC)}, pages 275--280, 2004.

\bibitem{Jimborean14}
A.~Jimborean, K.~Koukos, V.~Spiliopoulos, D.~Black-Schaffer, and S.~Kaxiras.
\newblock Fix the code. don't tweak the hardware: A new compiler approach to
  voltage-frequency scaling.
\newblock In {\em Proceedings of Annual IEEE/ACM International Symposium on
  Code Generation and Optimization}, CGO '14, pages 262--272, New York, NY,
  USA, 2014. ACM.

\bibitem{Kamruzzaman11}
M.~Kamruzzaman, S.~Swanson, and D.~M. Tullsen.
\newblock Inter-core prefetching for multicore processors using migrating
  helper threads.
\newblock In {\em Proceedings of the sixteenth international conference on
  Architectural support for programming languages and operating systems},
  ASPLOS '11, pages 393--404, New York, NY, USA, 2011. ACM.

\bibitem{koukos13}
K.~Koukos, D.~Black-Schaffer, V.~Spiliopoulos, and S.~Kaxiras.
\newblock Towards more efficient execution: A decoupled access-execute
  approach.
\newblock In {\em Proceedings of the 27th International ACM Conference on
  International Conference on Supercomputing}, ICS '13, pages 253--262, New
  York, NY, USA, 2013. ACM.

\bibitem{Liao06}
W.~Liao, L.~He, and K.~M. Lepak.
\newblock Temperature and supply voltage aware performance and power modeling
  at microarchitecture level.
\newblock {\em Trans. Comp.-Aided Des. Integ. Cir. Sys.}, 24(7):1042--1053,
  Nov. 2006.

\bibitem{Liu07thermalvs}
Y.~Liu, H.~Yang, R.~P. Dick, H.~Wang, and L.~Shang.
\newblock Thermal vs energy optimization for dvfs-enabled processors in
  embedded systems.
\newblock In {\em International Symposium on Quality Electronic Design
  (ISQED)}, pages 204 -- 209, 2007.

\bibitem{LLVM}
The llvm compiler infrastructure.
\newblock \url{http://llvm.org}.

\bibitem{Martin02}
S.~M. Martin, K.~Flautner, T.~Mudge, and D.~Blaauw.
\newblock Combined dynamic voltage scaling and adaptive body biasing for lower
  power microprocessors under dynamic workloads.
\newblock In {\em IEEE/ACM International Conference on Computer-aided Design
  (ICCAD)}, pages 721--725, 2002.

\bibitem{Mittal87}
S.~Mittal and Z.~Zhang.
\newblock Encache: A dynamic profiling based reconfiguration technique for
  improving cache energy efficiency.
\newblock {\em Journal of Circuits, Systems, and Computers (JCSC)}, 23, 2014.

\bibitem{papi99}
P.~J. Mucci, S.~Browne, C.~Deane, and G.~Ho.
\newblock Papi: A portable interface to hardware performance counters.
\newblock In {\em Proceedings of the Department of Defense HPCMP Users Group
  Conference}, pages 7--10, 1999.

\bibitem{mitosis05}
C.~G. Qui\~{n}ones, C.~Madriles, J.~S\'{a}nchez, P.~Marcuello, A.~Gonz\'{a}lez,
  and D.~M. Tullsen.
\newblock Mitosis compiler: An infrastructure for speculative threading based
  on pre-computation slices.
\newblock In {\em Proceedings of the 2005 ACM SIGPLAN Conference on Programming
  Language Design and Implementation}, PLDI '05, pages 269--279, New York, NY,
  USA, 2005. ACM.

\bibitem{Rauchwerger95}
L.~Rauchwerger, N.~M. Amato, and D.~A. Padua.
\newblock Run-time methods for parallelizing partially parallel loops.
\newblock In {\em Proceedings of the 9th International Conference on
  Supercomputing}, ICS '95, pages 137--146, New York, NY, USA, 1995. ACM.

\bibitem{Salz91}
J.~Saltz, R.~Mirchandaney, and K.~Crowley.
\newblock Run-time parallelization and scheduling of loops.
\newblock {\em Computers, IEEE Transactions on}, 40(5):603--612, May 1991.

\bibitem{Sembrant12}
A.~Sembrant, D.~Black-Schaffer, and E.~Hagersten.
\newblock Phase guided profiling for fast cache modeling.
\newblock In {\em IEEE / ACM Int'l Symp. on Code Generation and Optimization
  (CGO)}, pages 175--185, 2012.

\bibitem{pows}
V.~Spiliopoulos, A.~Sembrant, and S.~Kaxiras.
\newblock Power-sleuth: A tool for investigating your program's power
  behaviour.
\newblock {\em 2012 IEEE 20th International Symposium on Modeling, Analysis \&
  Simulation of Computer and Telecommunication Systems}, 0:241--250, 2012.

\bibitem{javajit}
T.~Suganuma, T.~Yasue, M.~Kawahito, H.~Komatsu, and T.~Nakatani.
\newblock A dynamic optimization framework for a java just-in-time compiler.
\newblock {\em SIGPLAN Not.}, 36:180--195, 2001.

\bibitem{Weiser81}
M.~Weiser.
\newblock Program slicing.
\newblock In {\em Int'l Conf. on Software Engineering (ICSE)}, pages 439--449,
  1981.

\bibitem{Yokota02}
D.~Yokota, S.~Chiba, and K.~Itano.
\newblock A new optimization technique for the inspector-executor method.
\newblock In {\em Proc. of the International Conference on Parallel and
  Distributed Computing Systems (PDCS)}, pages 706--711, 2002.

\bibitem{Zhang07}
W.~Zhang, D.~Tullsen, and B.~Calder.
\newblock Accelerating and adapting precomputation threads for effcient
  prefetching.
\newblock In {\em Int'l Symp. on High-Performance Computer Architecture
  (HPCA)}, pages 85--95, 2007.

\end{thebibliography}

\end{document}